\documentstyle[11pt]{article}

\def\dalemb#1#2{{\vbox{\hrule height .#2pt
        \hbox{\vrule width.#2pt height#1pt \kern#1pt
                \vrule width.#2pt}
        \hrule height.#2pt}}}

\def\td{\tilde}

\let\a=\alpha

\def\nn{\nonumber} \def\bd{\begin{document}} \def\ed{\end{document}}
\def\ds{\documentstyle} \let\fr=\frac \let\bl=\bigl \let\br=\bigr
\let\Br=\Bigr \let\Bl=\Bigl
\let\bm=\bibitem
\let\na=\nabla
\let\pa=\partial \let\ov=\overline
\newcommand{\be}{\begin{equation}}
\newcommand{\ee}{\end{equation}}
\def\ba{\begin{array}}
\def\ea{\end{array}}
\def\ft#1#2{{\textstyle{{\scriptstyle #1}\over {\scriptstyle #2}}}}
\def\fft#1#2{{#1 \over #2}}
\def\del{\partial}
\def\ie{{\it i.e.\ }}
\def\sst#1{{\scriptscriptstyle #1}}
\def\oneone{\rlap 1\mkern4mu{\rm l}}
\newcommand{\ho}[1]{$\, ^{#1}$}
\newcommand{\hoch}[1]{$\, ^{#1}$}
\newcommand{\bea}{\begin{eqnarray}}
\newcommand{\eea}{\end{eqnarray}}
\newcommand{\ra}{\rightarrow}
\newcommand{\lra}{\longrightarrow}
\newcommand{\Lra}{\Leftrightarrow}
\newcommand{\ap}{\alpha^\prime}
\newcommand{\bp}{\tilde \beta^\prime}
\newcommand{\tr}{{\rm tr} }
\newcommand{\Tr}{{\rm Tr} }
\newcommand{\NP}{Nucl. Phys. }
\newcommand{\tamphys}{\it Center for Theoretical Physics,
Texas A\&M University, College Station, Texas 77843}

\newcommand{\auth}{H. L\"u and C.N. Pope}

\begin{document}

\hfill{CTP TAMU-56/96}

\hfill{hep-th/9611079}

\vspace{20pt}

\begin{center}
{ \large {\bf Domain Walls from M-branes}}

\vspace{30pt}
\auth

\vspace{15pt}
{\tamphys}
\vspace{40pt}

\underline{ABSTRACT}
\end{center}
\vspace{40pt}

     We discuss the vertical dimensional reduction of M-branes to domain
walls in $D=7$ and $D=4$, by dimensional reduction on Ricci-flat 4-manifolds
and 7-manifolds.  In order to interpret the vertically-reduced 5-brane as a
domain wall solution of a dimensionally-reduced theory in $D=7$, it is
necessary to generalise the usual Kaluza-Klein ansatz, so that the 3-form
potential in $D=11$ has an additional term that can generate the necessary
cosmological term in $D=7$. We show how this can be done for general
4-manifolds, extending previous results for toroidal compactifications.  By
contrast, no generalisation of the Kaluza-Klein ansatz is necessary for the
compactification of M-theory to a $D=4$ theory that admits the domain wall
solution coming from the membrane in $D=11$. 

{\vfill\leftline{}\vfill
\vskip	10pt
\footnoterule
{\footnotesize	Research supported in part by DOE
Grant DE-FG05-91-ER40633 \vskip	-12pt}}

\pagebreak
\setcounter{page}{1}

      It has recently been observed that some unusual and interesting 
features emerge when one pushes the process of vertical dimensional 
reduction of $p$-brane solutions to the stage where the transverse space 
orthogonal to the $d=p+1$ dimensional world-volume becomes only one
dimensional.  More precisely, we may begin with extremal $p$-brane solutions
in $D$-dimensional supergravity, which are described in terms of an harmonic
function $H$ on the $(D-p-1)$-dimensional transverse space \cite{lpss1}: 
\bea
ds_{\sst D}^2 &=& H^{-\ft{4\td d}{\Delta (D-2)}}\, dx^\mu dx^\nu \, 
\eta_{\nu\mu} + H^{\ft{4d}{\Delta(D-2)}}\, dy^m\, dy^m\ ,\nn\\
e^{\phi} &=& H^{\ft{2 a}{\epsilon\Delta}} \ ,\label{dsol}
\eea
where $\del_m \del_m H=0$, $\epsilon=\pm1$ according to whether the solution 
is elementary or solitonic, $\td d=D-d-2$, and the constants $a$ and
$\Delta$, related by $a^2=\Delta -2d\td d/(D-2)$, characterise the dilaton
coupling in the Lagrangian 
\be
e^{-1}\, {\cal L} = R -\ft12(\del\phi)^2 -\fft{1}{2 n!}\, e^{a\phi} \,
F_n^2\ .
\ee
The $n$-rank field strength $F_n$ has $n=p+2$ in the elementary case, or
$n=(D-p-2)$ in the solitonic case.  In the $p$-brane solution, it is given by
\bea
{\rm elementary}: && F_{m\mu_1\cdots\mu_n}= \epsilon_{\mu_1\cdots\mu_n}\,
\del_m H\ ,\nn\\
{\rm solitonic}: && F_{m_1\cdots m_n} = \epsilon_{m_1\cdots m_n\ell}\, 
\del_\ell H\ .\label{fsol}
\eea
By stacking up an infinite periodic 
array of single $p$-brane solutions along an axis in the transverse space, 
one can construct a solution which, in the continuum limit, becomes 
independent of the associated coordinate.  The solution may then be 
dimensionally reduced along this coordinate, to give a $p$-brane in $(D-1)$
dimensions.  It is described in terms of a function that is harmonic on the 
dimensionally-reduce transverse space. The process may be continued until 
eventually the $p$-brane is living in a spacetime whose total dimension is 
only $(p+2)$.  The solution describes a domain wall, and has the form given 
by (\ref{dsol}) and (\ref{fsol}), with $H= 1 + m|y|$.  

    The above discussion of the dimensional reduction of the $p$-brane
solutions has a parallel at the level of the theory itself, namely that all
the dimensionally-reduced $p$-branes are solutions of the
dimensionally-reduced Lagrangian. This is straightforward to understand until
the vertical descent reaches the domain wall discussed above.  At this 
point, a significant difference between the elementary and solitonic 
solutions emerges.  In fact the elementary case continues to follow the
standard pattern, but in the solitonic case, the domain-wall solution
requires the presence of a cosmological constant term in the
$(p+2)$-dimensional Lagrangian, of the form $m^2\, e\, e^{a\phi}$. However,
such a cosmological term is not generated in the standard Kaluza-Klein
reduction procedure.  The resolution of this puzzle is that a slight
generalisation of the standard Kaluza-Klein reduction is needed here, and in
fact, the form of the domain-wall solution already indicates the nature of
this generalisation.  We can see from the expression (\ref{fsol}) for the
solitonic 1-form field strength in the $(p+3)$-dimensional solution that we
have $F= m dz$, where $z$ is the additional transverse-space coordinate that
will be compactified in the final reduction step that gives the domain wall.
It follows that the 0-form potential (or axion) for this field strength
must be of the form $A_0=m z$.  Thus the axion depends on the coordinate $z$
that is to be compactified.  

     Normally in the Kaluza-Klein procedure, one performs a truncation in
which all the fields are taken to be independent of the compactification
coordinate.  By making this requirement, one ensures that the truncation is
consistent, {\it i.e.\ }that all the solutions of the lower-dimensional
theory are also solutions of the higher-dimensional one. However, this
$z$-independent truncation is slightly more restrictive than is actually
necessary.  For consistency, one need require only that the
higher-dimensional equations of motion, after substituting the Kaluza-Klein
ansatz, should be independent of $z$. In particular, this means that a
potential such as $A_0$ above can be allowed to have a linear dependence on
$z$, provided that it always appears {\it via} its exterior derivative $d
A_0$.  This generalisation of the ansatz gives rise to a cosmological term
in the lower-dimensional Lagrangian, since it implies that the reduction of 
the 1-form field strength now yields a ``0-form field strength'' as well as a
1-form.  It also gives rise to mass terms for
certain of the previously-massless gauge fields, and thus we may describe
the resulting theory as a massive supergravity.  This type of generalised
Kaluza-Klein dimensional reduction was first considered in \cite{bdgpt}
where the axion in type IIB theory in $D=10$ was taken to have an additional
term that is linearly dependent on the compactifying circular coordinate.
This gives rise to a maximally supersymmetric massive supergravity in $D=9$.
This procedure was generalised in \cite{clpst} to compactify M-theory to
various massive supergravities in lower dimensions.  (In fact generalised 
Kaluza-Klein ans\"atze that give rise to cosmological terms were also 
discussed in a general group theoretic framework in \cite{ss}.  It was also
observed, in the context of compactifying the heterotic string to $D=4$,
that wrapping the 5-brane on a 3-torus to give rise to a membrane in $D=4$
would require some ansatz that went beyond the usual Kaluza-Klein
dimensional reduction \cite{dk,dkmr}.) The occurrence of the cosmological
term explains how the domain-wall, which comes from a solution in the higher
dimension, can continue to be a solution in this lower dimensional massive
theory.  It should remarked, however, that just as the domain wall is a
solution of the massive theory but not the massless one, so $p$-branes that
are solutions of the massless theory will not be solutions of the massive
one. 

    The generalised Kaluza-Klein reduction described above entails making 
the ansatz
\be
A_0(x,z) = m \, z + A_0(x)\label{gkk}
\ee
for the axion $A_0$, while making the standard $z$-independent ansatz for 
all the other fields.  We may view $z$ as a 0-form defined locally on the 
$S^1$ compactification manifold, whose exterior derivative gives the volume 
form $dz$ on $S^1$.  This formulation lends itself to an immediate 
generalisation, in which we consider a Kaluza-Klein compactification on an 
$n$-dimensional manifold $M_n$, with a generalised ansatz for an 
$(n-1)$-form potential $A_{n-1}$, of the form
\be
A_{n-1}(x,z) = m\, \omega_{n-1}(z) + A_{n-1}(x) + A_{n-2}^\a(x)\wedge 
\Omega_1^\a(z) + \cdots \ ,\label{gkk2}
\ee
where $x$ denotes the lower-dimensional coordinates, $z$ denotes the 
coordinates on the internal manifold $M_n$, $\Omega_q^\a$ denotes the set of 
harmonic $q$-forms on $M_n$, and $\omega_{n-1}$ is an $(n-1)$ form defined
locally on $M_n$, whose exterior derivative gives the globally-defined
volume form: $d\omega_{n-1} = \Omega_n$.  Provided that $A_{n-1}$ appears
only {\it via} its exterior derivative in the higher-dimensional equations
of motion, the inclusion of the extra term $m\, \omega_{n-1}$ in its ansatz
will not upset the consistency of the Kaluza-Klein truncation.\footnote{The
consistency of the truncation even without the inclusion of the extra term
is far from obvious in generic compactifications.   The $S^1$
compactification is rather special in that the truncation to the
$z$-independent harmonics on $S^1$ is guaranteed to be consistent, since one
is setting all the non-singlets under the $U(1)$ isometry group to zero, and
so the retained singlet modes cannot, {\it via} the non-linear terms in the
field equations, generate sources for the non-singlet modes in the
higher-dimensional equations of motion.  By contrast, in a compactification
on a manifold $M_n$ such as K3, Calabi-Yau, or a Joyce manifold, the
non-linear terms involving products of harmonic forms on $M_n$ would in
general be expected to generate higher harmonics that would act as sources
for fields that have been set to zero in the truncation. Indeed this would
be the case for such a compactification of a generic theory, and it  seems
that it is supersymmetry that forbids the appearance of these source terms,
and thus ensures the consistency of the truncation \cite{ps,dfps}, in
supergravity theories.}  As in the case of the 0-form example that we
discussed previously, the dimensional reduction of kinetic term for the
field strength $F_n=dA_{n-1} +\cdots$ will now generate a cosmological term
in the lower-dimensional theory, which consequently will admit domain-wall
solutions. 

     In this letter, we shall concentrate on an example of particular 
interest, namely compactifications of $D=11$ supergravity, or M-theory.  
Since this has just a 4-form field strength, it implies that we can consider 
either generalised Kaluza-Klein reductions of the above kind on 4-manifolds,
or else standard Kaluza-Klein reductions on 7-manifolds, depending on
whether we make a solitonic ansatz or an elementary ansatz.  The
corresponding M-branes, \ie a 5-brane or a membrane, vertically reduce to
domain-wall solutions in 7 or 4 dimensions respectively. 

     First we shall look at the solitonic ansatz for the 4-form field
strength.  This will illustrate the new features arising from the need for 
the generalisation of the Kaluza-Klein reduction procedure. The simplest
choice for the internal 4-manifold $M_4$ is the 4-torus.  This case has in
fact already been discussed in \cite{clpst}, from a slightly different
viewpoint in which the $D=11$ theory is first compactified on a 3-torus,
giving rise to a set of field strengths in $D=8$ that includes a 1-form
coming from the 4-form of $D=11$.  Then, its associated axionic potential
$A_0$ is subjected to the generalised Kaluza-Klein reduction of the form
(\ref{gkk}), generating the massive theory in $D=7$. In our present
discussion, the same result is achieved in one step, by making the
generalised ansatz 
\bea
A_3(x,z) &=& m \, z_4\, dz_1\wedge dz_2 \wedge dz_3 + A_3(x)\nonumber\\
&& + A_2^i(x)\wedge dz_i  +\ft12 A_1^{ij}(x)\wedge dz_i\wedge dz_j + \ft16
A_0^{ijk}(x)\wedge dz_i\wedge dz_j \wedge dz_k\label{4torus} 
\eea
in $D=11$, rather than $A_0(x,z_4) = m\, z_4 + A_0(x)$ in $D=8$.  As has
been shown in \cite{clpst}, the resulting theory in $D=7$ has a topological
mass term for $A_3$, ordinary mass terms for the four Kaluza-Klein vectors,
and a cosmological term.  The remaining six 1-form potentials, and four
2-form potentials, which come from $A_3$ in $D=11$, are massless, as are the
six 0-form potentials coming from the metric.  The four 0-form potentials
coming from $A_3$ in $D=11$ are eaten when the four Kaluza-Klein vectors
become massive. 

     This massive theory in $D=7$ is maximally supersymmetric; namely it has
the same number $N=2$ of supersymmetries as the usual massless $D=7$ maximal
supergravity. However, unlike a normal Poincar\'e or de Sitter supergravity,
it does not admit any solution, such as Minkowski or anti-de Sitter
spacetime, that preserves all the supersymmetry. In fact, its analogous 
``natural'' ground state is the domain wall solution that we mentioned 
previously.  This can be seen at the level of the dimensional reduction of 
solutions by noting that in the above generalised reduction, the 5-brane
solution in $D=11$ is vertically reduced to a domain wall in $D=7$.  The
metric in $D=11$ is given by (\ref{dsol}), with $H$ chosen to be harmonic in
just a 1-dimensional subspace of the 5-dimensional transverse space, \ie 
\be
ds_{11}^2 = (1+ m\, |y|)^{-\ft13}\, dx^\mu dx^\nu\, \eta_{\mu\nu} +  
(1+m\, |y|)^{\ft23}\, (dy^2 + d\bar s_4^2)\ ,\label{sol1}
\ee
where $d\bar s_4^2$ is the metric on the compactifying 4-torus.  The 4-form 
field strength is given by $F_4= m\, \Omega_4 = m\, dz_1\wedge dz_2 \wedge 
dz_3 \wedge dz_4$, which is consistent with the ansatz (\ref{gkk2}) for the 
generalised dimensional reduction of the 3-form potential.  The domain wall 
preserves one half of the supersymmetry of the $D=7$ massive supergravity.

     We can now easily extend the above discussion to the case where $d\bar 
s_4^2$ is the metric on any Ricci-flat compactifying 4-manifold.  The metric 
(\ref{sol1}) remains a solution for the 11-dimensional supergravity 
\cite{dlps}.  We shall consider the example of a K3 compactification in some
detail. Clearly the solution (\ref{sol1}) again describes a domain wall
after reduction down to $D=7$.  Our principal interest is to investigate the
structure of the supergravity theory in $D=7$ that we obtain by making the
generalised dimensional reduction on K3.  As in the previous toroidal
compactification, so also in this case the generalised dimensional reduction
on K3 will give a theory with the same supersymmetry as that given in a
standard Kaluza-Klein reduction on K3.  Thus in this
case, we will obtain a massive $D=7$ theory with $N=1$ supersymmetry.  The
analogue of the generalised ansatz (\ref{4torus}) in this case will be 
\be
A_3(x,z)= m\, \omega_3 + A_3(x)+A_1^\a(x)\wedge \Omega_2^\a(z)\ ,\label{3form}
\ee 
where $d\omega_3=\Omega_4$ is the volume form of K3, and the summation in
the final term is over the 22 harmonic 2-forms $\Omega_2^\a$ on K3.  Note
that in the K3 compactification, there are no 1-form potentials coming from
the metric, since there are no harmonic 1-forms on K3, unlike on the
4-torus.  There are, however, a total of 58 massless scalars coming from the
metric, corresponding to the 57 varieties of volume-preserving Ricci-flat
deformations, together with the overall scale deformation.  Upon
substitution of the ans\"atze into the $D=11$ Lagrangian, we find that the
kinetic term for $F_4$ reduces to terms including a cosmological term, while
the $F_4\wedge F_4\wedge A_3$ term in $D=11$ gives rise to a topological
mass term for $A_3$ in $D=7$.  The $D=7$ Lagrangian has the form 
\bea
{\cal L} &=& e\, R -\ft12e\, (\del\phi)^2 -\ft12  m^2\, e\, 
e^{\ft{8}{\sqrt{10}}\, \phi} -\ft1{48}e\, e^{-\ft{4}{\sqrt{10}}\phi}\, F_4^2
\nn\\
&&+\ft1{288}m\, \epsilon^{\sst{MNPQRST}}\, F_{\sst{MNPQ}}\, 
A_{\sst{RST}} + {\cal L}_{\rm matter}\ ,\label{k3lag}
\eea
where $\phi$ is the scalar field associated with the scaling mode of K3, and 
${\cal L}_{\rm matter}$ represents the Lagrangian for the 57 scalars and 22 
vectors of the matter multiplets, and their couplings to the supergravity 
multiplet.  If we truncate out the matter multiplets, the resulting $N=1$ 
pure massive supergravity presumably coincides with the topologically 
massive theory obtained in \cite{mtn}.

     It is of interest to see what happens if we take the above 
topologically massive $D=7$, $N=1$ supergravity plus matter, and compactify
it by one further dimension, using the standard Kaluza-Klein ansatz on a
circle.  The resulting theory in $D=6$ has a cosmological term, and an
off-diagonal mass term involving a 2-form and a 3-form field strength.  This
theory is dual to the one obtained by performing a K3
compactification of the type IIA theory in $D=10$, where the 3-form
potential takes the generalise form (\ref{3form}).  Put another way, this
means that an ordinary $S^1$ reduction of M-theory followed by a generalised
K3 reduction gives a theory in $D=6$ that is equal to the one obtained by
first making a generalised K3 reduction of M-theory, followed by an ordinary
$S^1$ reduction.   Although this would obviously be true if the reductions
were all of the standard Kaluza-Klein kind, this is not as trivial as it
sounds in the present case, since if instead of K3 we used $T^4$ for the
generalised reduction, then the order in which the ordinary and the
generalised reduction steps are performed matters.  In fact in general,
reductions of the ordinary and the generalised forms do not commute.  This
was observed in the case of $S^1$ reductions in \cite{clpst}. 

     Now we shall consider the situation where the 4-form field strength
carries an electric charge instead, giving a membrane solution and an
8-dimensional transverse space.  We wish to perform a vertical dimensional
reduction to a domain wall in $D=4$, by taking the harmonic function
governing the solution in $D=11$ to depend on only one of the eight transverse
coordinates, giving
\be
ds_{11}^2 = (1+ m|y|)^{-\ft23}\, dx^\mu dx^\nu\, \eta_{\mu\nu} +
(1+m|y|)^{\ft13}\, (dy^2 + d\bar s_7^2)\ .\label{membrane}
\ee
In this electric case, the potential $A_3$ for the membrane 
solution in $D=11$ is given by $A_3= H\, \epsilon_3$, where $\epsilon_3 =
dx^0\wedge dx^1 \wedge dx^2$ is the volume form on the membrane world 
volume, and $H=1+m|y|$.  Thus unlike the vertical reduction of the 5-brane 
that we discussed previously, here the 3-form potential is independent of 
the compactifying coordinates of the 7-metric $d\bar s_7^2$. 
Correspondingly, it is not obligatory in this case to consider a generalised 
Kaluza-Klein ansatz for the dimensional reduction of the $D=11$ theory to 
$D=4$ in order to obtain a domain wall solution.\footnote{Generalised 
reductions of M-theory to $D=4$ will be discussed in \cite{llp}.}  The 
reason why the standard Kaluza-Klein reduction is able to give a 
theory that admits domain wall solutions is that now we will end up in $D=4$ 
with a theory that includes a 4-form field strength.  The relevant terms in 
the $D=4$ Lagrangian are of the form
\be
{\cal L} = e R -\ft12 e\, (\del\phi)^2 -\ft1{48} \, e\, e^{a\phi} F_4^2\ .
\label{f4lag}
\ee
This admits an electric membrane solution (\ie a domain wall) of the 
standard $p$-brane type.  Note that one could dualise $F_4$ to a ``0-form 
field strength,'' or cosmological term, giving the Lagrangian 
\be
{\cal L} = e R -\ft12 e\, (\del\phi)^2 -\ft1{2} \, e\, m^2\, e^{-a\phi}
\ ,\label{f0lag}
\ee
where $m$ is the constant of integration coming from solving the equation of 
motion for $F_4$.  We can recognise this dualised form of the Lagrangian as 
being of the same kind, with a cosmological term, as the $D=7$ Lagrangian 
that we obtained earlier by generalised Kaluza-Klein reduction.  Indeed, it 
was observed in \cite{bdgpt} that an alternative way of understanding the 
generalised reduction procedure was by first dualising the 1-form field 
strength in $(D+1)$ dimensions to a $D$-form field strength, then performing 
a standard Kaluza-Klein eduction to $D$ dimensions, and finally dualising 
the resulting $D$-form field strength to a 0-form field strength, giving the 
cosmological term.  It should be emphasised, however, that the Lagrangian 
(\ref{f0lag}) is not completely equivalent to (\ref{f4lag}): in the former, 
since $m$ is a given fixed parameter in the Lagrangian, it does not admit a 
Minkowski ground state if $m$ is non-zero, and conversely, it does not
admit the domain-wall ground state if $m$ is zero.  On the other hand, the 
Lagrangian (\ref{f4lag}) admits both the Minkowski and the domain wall 
ground states.

     The seven-dimensional internal metric $d\bar s_7^2$ may be taken to be 
any Ricci-flat metric.  The simplest choice is the 7-torus, which was 
discussed in \cite{clpst}, but we may instead consider other possibilities, 
such as K3$\times T^3$ (which was discussed in the context of the Minkowski 
vacuum state in \cite{dnp}), Y$\times S^1$, where Y is any 6-dimensional 
Calabi-Yau space, and J, where J is any 7-dimensional Joyce manifold.
These last cases are compact Ricci-flat manifolds with $G_2$ holonomy
\cite{j}.  The K3$\times T^3$, Y$\times S^1$ and J compactifications will
give respectively $N=4$, $N=2$ and $N=1$ supergravity in $D=4$.  By
including the 4-form field strength in $D=4$, all these supergravity
theories admit domain wall ground states as well as Minkowski ground states.
 The domain wall ground states are the vertical dimensional reductions of
the membrane in $D=11$. 

     The counting of fields in $D=4$, determined from the Betti numbers of 
the compactifying manifolds, is as follows.  For the K3$\times T^3$ 
compactification, we obtain 28 2-form field strengths and 134 scalars 
\cite{dnp}, together with the 4-form field strength that can support the domain 
wall solution.  For a Y$\times S^1$ compactification, we get $(b_2+1)$ 2-form 
field strengths, $(2 b_3 +b_2 +1)$ scalars and the 4-form, where $b_2$ and 
$b_3$ are the Betti numbers of the Calabi-Yau manifold Y.  For the J 
compactifications, we get $b_2$ 2-form field strengths, $2 b_3$ scalars, and 
the 4-form, where $b_2$ and $b_3$ are the Betti numbers of the Joyce
manifold J.  In counting the scalars in this last example, we have made use
of results in \cite{gpp} for the counting of Lichnerowicz zero modes on
manifolds of exceptional holonomy. 

     In summary, in this letter we have studied the vertical dimensional
reduction of $p$-brane solutions to domain walls in $D=p+2$.  In particular, 
we focussed on the domain wall solutions in 7 and 4 dimensions which are the 
vertical reductions of the solitonic 5-brane and elementary membrane in
M-theory.  These reductions can be achieved by taking the harmonic function
governing the M-brane in $D=11$ to be independent of 4 or 7 transverse-space
directions respectively, and compactifying these directions on a Ricci-flat
4-manifold or 7-manifold.  We also studied how M-theory is dimensionally
reduced on these manifolds.  In order for the dimensionally-reduced theory 
in $D=7$ to be able to admit, as it must, the solitonic domain wall solution
coming from the 5-brane, it is necessary to generalise the usual 
Kaluza-Klein ansatz to allow the 3-form potential to have an additional term 
whose exterior derivative is a constant multiple of the volume form of the 
compactifying 4-manifold.  This does not spoil the consistency of the 
Kaluza-Klein truncation, since the 3-form potential always enters in the 
equations of motion {\it via} its exterior derivative.  This generalised 
dimensional reduction gives rise to a supergravity theory in $D=7$ with a
topological mass term and a cosmological term, together possibly with other 
mass terms depending on the choice of compactifying 4-manifold.  The 
supersymmetry of the massive theory also depends on the choice for the 
4-manifold; for example we get $N=2$ for the 4-torus, and $N=1$ for K3.

     The situation for the 4-dimensional theory obtained by compactification 
on a Ricci-flat 7-manifold is different.  In this case, the standard 
Kaluza-Klein ansatz for the dimensional reduction is adequate.  The reason
why there is nevertheless a domain wall solution is that there is a 4-form
field strength in $D=4$, which is dual to a cosmological term.  The major 
difference in this case is that the dimensionally-reduced theory admits 
both domain wall solutions and the usual $p$-brane solutions, as well as a 
Minkowski spacetime vacuum solution.  By contrast, in the generalised 
Kaluza-Klein reduction to $D=7$, the resulting massive theory is different 
from the usual massless theory obtained by the standard reduction; the 
former admits only the domain wall solution, while the latter admits only 
the usual $p$-brane solutions and the Minkowski vacuum.  Finally, we remark 
that yet more general compactifications utilising the cohomology classes of 
the internal manifold are possible.  These will be the subject of a 
forthcoming publication \cite{llp}.

\section*{Acknowledgement}

      We are grateful to M.J. Duff, I.V. Lavrinenko, K.S. Stelle and P.K.
Townsend for useful discussions. 


\end{document}